\documentclass[12pt]{article}

\usepackage{amsmath}
\usepackage{amssymb}
\usepackage{bm}
\usepackage{graphicx}
\usepackage{enumerate}
\usepackage{natbib}
\usepackage{hyperref}
\usepackage{url} 
\usepackage{caption}
\usepackage{subcaption}
\usepackage{tikz}
\usepackage{tikz-network}
\usetikzlibrary{calc}
\usepackage{physics}
\usepackage{bbm}
\usepackage{easyReview}
\DeclareMathOperator{\EX}{\mathbb{E}} % Expected value

\newcommand{\Mult}{\text{Multinom}} % Multinomial distribution

\begin{document}

\title{Causal drivers of dynamic networks}
\author{%
Melania Lembo$^{a \footnote{Corresponding author: melania.lembo@usi.ch.}}$, Ester Riccardi$^b \footnote{ester.ricc@gmail.com}$, Veronica Vinciotti$^b \footnote{veronica.vinciotti@unitn.it}$, Ernst C. Wit$^a \footnote{ernst.jan.camiel.wit@usi.ch}$ \\ \\
	\small{$a.$ Institute of Computing, Universit\`a della Svizzera italiana} \\
  \small{$b.$ Department of Mathematics, University of Trento} \\
}
\date{}

\maketitle

\begin{abstract}%
Dynamic networks models describe temporal interactions between social actors, and as such have been used to describe financial fraudulent transactions, dispersion of destructive invasive species across the globe, and the spread of fake news. An important question in all of these examples is what are the causal drivers underlying these processes. Current network models are exclusively descriptive and based on correlative structures.

In this paper we propose a causal extension of dynamic network modelling. In particular, we prove that the causal model satisfies a set of population conditions that uniquely identifies the causal drivers. The empirical analogue of these conditions provide a consistent causal discovery algorithm, which distinguishes it from other inferential approaches. Crucially, data from a single environment is sufficient. We apply the method in an analysis of bike sharing data in Washington D.C. in July 2023.      
\end{abstract}

\section{Introduction}
In this manuscript, our aim is to define a causal dynamic network by means of a structural relational event model. Relational event models were first introduced in \cite{butts2008} to study instantaneous interactions. Relational events are defined as discrete events in which a sender initiates an interactions directed towards one or more receivers. \cite{bianchi2023} provide an extensive overview of the developments over the past 15 years of these type of dynamic network models. 

Causal discovery aims to find causal relationships between two or more variables, trying to identify which variables affect the outcome. Significant progress has been made in this area over the past 30 years. Indeed, several techniques for detecting causal drivers of a target variable have been developed. First described in \cite{spirtes1991}, the PC algorithm aims to identify the Markov equivalence class of the true causal graph using only conditional independence tests. The PC algorithm comes with some drawbacks. First of all, the complexity increases exponentially with the number of nodes of the graph. Secondly, it is not always possible to define the orientation of all the edges. Indeed, the result is a large class of equivalent DAGs. 

The instrumental variable (IV) method identifies the causal structure of a target of interest via an endogenous variable and some conditional independence assumptions associated with this variable \citep{angrist1996}. It aims to resolve the problem of the presence of unobserved confounders that may affect both the treatment and the outcome.  

Recently, new ideas of causal discovery have been suggested that exploit the property of a causal model to be invariant under intervention on variables other than the target one. \cite{peters2016} show that in a linear Structural Equation Model (SEM) with no confounders invariant causal prediction holds, meaning that under any arbitrary distribution assigned to the covariates $X$ both the linear predictor and the distribution of the additive noise of the target $Y$ are constant. Unfortunately, this method requires that data must be accessible across a number of sufficiently different environments. Furthermore,  the procedure is computationally expensive, since it involves a large number of independence statistical tests to identify the causal model. \cite{rothenhäusler2019} build forth on the invariance idea, by proposing a strategy based on inner-product invariance of the causal residual and the covariates. In principle, observations from the observational environment together with data from a sufficiently richly perturbed environment is sufficient to make the causal model identifiable.

In this paper we will consider a dynamic network $N =\{N_{sr}(t)~|~N_{sr}(t)\in \mathbb{N}_0\}$ and a multivariate covariate process $\boldsymbol{X}=\{X_{sr}(t)~|~X_{sr}(t)\in \mathbb{R}^p\}$, where the subscripts $(s,r)$ refer to the directed edges of the network. The main question we are interested in answering is which of the covariates $\boldsymbol{X}$ causally affect the dynamic network $N$. We will define $(X,N)$ as a structural equation model. Using ideas from invariant causal prediction and focusing on a target variable of interest $N$, its causal parents $X_{PA}$ affect its distribution and therefore satisfy certain invariance properties, whereas its causal children $X_{CH}$ are consequences of the target itself, which violate these invariance properties. In section~\ref{sec:causalrem} we formally define the dynamic network model as a counting process. Under the weak assumption of no simultaneous links, the causal likelihood of the network process can be shown to satisfy a crucial invariance property. Although by itself it is not sufficient to identify the causal model, we show that together the standard MLE property of the causal model, it does identify the causal model uniquely (up to a zero set). In Section~\ref{sec:sim} we show the performance of the method in a simulation study, whereas in section~\ref{sec:bike} we illustrate the method by means of a dataset involving bike sharing dynamic network dataset.

\section{Causal Dynamic Networks Models}
\label{sec:causalrem}

% \textcolor{red}{Very briefly go into what the options in terms of dynamic network modelling (???) are and directly focus on REMs. Counting process, partial likelihood and NCC to get to the logistic likelihood. Technically we use global covariates too in the bike sharing analysis but probably good to put nothing in here on it.}

In this section, we define a causal dynamic network model by means of a structural relational event model. This model describes the causal structure of temporal interactions between vertices of a one-mode or two-mode network. The temporal causal structure should not be confused with Granger causality. Instead, the causal nature of the network is defined as a risk invariance under arbitrary perturbations of covariates.

\subsection{Structural Relational Event  Model}
In this section we define the structural relational event model as the joint temporal process of a dynamic network $N$ and a set of potential causal drivers $\boldsymbol{X}$ on some temporal interval $[0,T]$, i.e., $\{(N(t),\boldsymbol{X}(t))\}_{t\in [0,T]}$, that satisfies some causal conditions. Let $V_1$ and $V_2$ be two sets of vertices. In the case of a one-mode dynamic network, we have that $V_1\equiv V_2$. A relational event $e$ is defined as a triplet of the type $(t,s,r)$, where $t \in [0,T]$ is the time of the event connecting, $s \in V_1$ is the sender of the event, and $r \in V_2$ is the receiver. Furthermore, we consider the existence of a marked point process $\mathcal{M} = \{(t_i,(s_i,r_i)):i \ge 1\}$, in which we assume unique event times $t_i$ and the interactions $(s_i,r_i)$ as marks, as well as a p-dimensional covariate process $\boldsymbol{X}=\{ \boldsymbol{X}_{sr}(t) \in \mathbb{R}^p~|~t\in [0,T], s\in V_1, r\in V_2\}$. Covariates can be endogenous, when they summarize some part of the dynamic of the network, such as reciprocity. Exogenous covariates are temporal processes that capture other aspects external to the network process, such as the physical distance between the vertices. 

\begin{figure}[t]
    \centering
    \resizebox{4cm}{!}{
    \begin{tikzpicture}
        \Vertex[x = 2, y = 0, label = X_1, Math, size = 1, color = white, fontscale = 1.5]{X1}
        \Vertex[x = 2, y = -2, label = X_2, Math, size = 1, color = orange!30, fontscale = 1.5]{X2}
        \Vertex[x = 4, y = -2, label = X_3, Math, size = 1, color = orange!30, fontscale = 1.5]{X3}
        \Vertex[x = 0, y = -4, label = X_4, Math, size = 1, color = white, fontscale = 1.5]{X4}
        \Vertex[x = 2, y = -4, label = N, Math, size = 1, color = white, fontscale = 1.5]{N}
        \Vertex[x = 4, y = -4, label = X_5, Math, size = 1, color = orange!30, fontscale = 1.5]{X5}
        \Vertex[x = 2, y = -6, label = X_7, Math, size = 1, color = orange!30, fontscale = 1.5]{X7}
        \Vertex[x = 4, y = -6, label = X_8, Math, size = 1, color = orange!30, fontscale = 1.5]{X8}
        \Vertex[x = 0, y = -6, label = X_6, Math, size = 1, color = white, fontscale = 1.5]{X6}
        \Edge[Direct](X1)(X2)
        \Edge[Direct](X1)(X3)
        \Edge[Direct](X2)(X3)
        \Edge[Direct](X2)(X4)
        \Edge[Direct](X2)(N)
        \Edge[Direct](X3)(N)
        \Edge[Direct](N)(X7)
        \Edge[Direct](N)(X8)
        \Edge[Direct](X5)(X8)
        \Edge[Direct](X7)(X6)
    \end{tikzpicture}}
    \caption{Causal REM with 8 covariates with nodes in orange belonging to the Markov Blanket of the relational process $N$. The causal parents of $N$ are $X_2$ and $X_3$.}
    \label{fig:causalrem}
\end{figure}
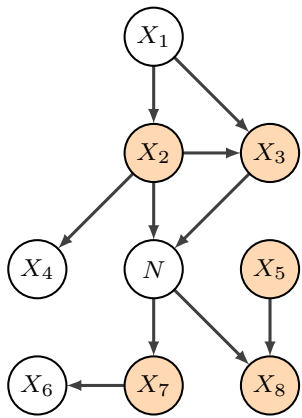

We define the structural relational event model as the joint process involving $\boldsymbol{X}$ and the multivariate counting process $N = \{N_{sr}\}_{(s,r)\,\in\,V_1\times V_2}$, associated with $\mathcal{M}$ that is structurally related to $\boldsymbol{X}$. In particular, each component $N_{sr}(t)$ records the number of interactions from $s$ to $r$ that occur until time $t$. Formally,
\begin{equation*}
    N_{sr}(t) = \sum_{i \ge 1} \mathbbm{1}_{\{ t_i \le t, s_i = s, r_i = r\}},
\end{equation*}
where $N_{sr}(0) = 0$. Under the above assumptions, the counting process is a submartingale. Hence, according to the Doob-Meyer decomposition theorem, it can be decomposed as follows:
\begin{equation*}
    N_{sr}(t) = \Lambda_{sr}(t) + M_{sr}(t),
\end{equation*}
where $\Lambda_{sr}(t)= \int_0^t \lambda_{sr}(\tau) \; d\tau$ represents the predictable part of the counting process and $M_{sr}(t)$ is a zero-mean martingale. From a structural point of view, the generative process $\Lambda$ only depends on $\boldsymbol{X}_{PA}$, where causal parents set $PA\subset \{1,\ldots,p\}$. In Figure~\ref{fig:causalrem}, the causal parents consist of covariates 2 and 3.

The hazard function $\lambda_{sr}(t)$ represents the rate of the event $(s,r)$ occurring at time $t$. Without loss of generality, the hazard rate can be written as
\begin{equation}
    \lambda_{sr}(t) = \mathbbm{1}_{\{(s,r)\in\mathcal{R}(t)\}} \, \lambda_0(t) \, \exp \left\{ f_{PA}(\boldsymbol{X}_{sr} (t))  \right\} . 
    \label{eq:hazard-rate}
\end{equation}
where the risk set $\mathcal{R}(t)$ is the set of events that are at risk of happening at time $t$, $\lambda_0 (t)$ is the baseline hazard function unrelated to $(s,r)$ and $f_{PA}: \mathbb{R}^p\rightarrow\mathbb{R}$ is the relative risk function depending only on the causal parents. A special case is where the function is linear, i.e., $f_{PA}(x) = \beta_{PA}^tx_{PA}$, where $\beta_{PA}$ is the linear causal parameter. 

The causal parameter solves a population likelihood maximization conditional on the true causal parents. Unfortunately, this property does not help to identify the causal parameters, as the overall maximization will in general identify a function that depends on all the covariates in the Markov blanket of $N$.

% E\left[\frac{1-\dot b(f_{PA}(X))}{\ddot b(f_{PA}(X))} \right]

\subsection{Invariant causal prediction for dynamic network models}
\label{subsec: icp_dyn_net}

The aim of this section is to define a property that identifies the causal risk function $f_{PA}$ irrespective of knowing the set of causal parents. We will define a partial likelihood for the process $N$ and show that its associated Pearson Risk is exactly 1 for the causal model,
\begin{eqnarray}
    R^P(f_{PA}) &=& 1. 
    \label{eq:pearson1}
\end{eqnarray}  
Together with population maximum likelihood the Pearson risk invariance property \eqref{eq:pearson1} can be shown to uniquely identify the structural relational event model. 

Let $n$ be the (random) number of relational events in $[0,T]$, we can define a nested case-control partial likelihood \citep{borgan1995,filippi2024stochastic} associated with the dynamic network $N$ as 
\begin{align*}
    \ell(f) &= \log \prod_{i=1}^n \frac{ \exp \left\{ f(\boldsymbol{X}_{s_i r_i} (t_i))  \right\} }{\exp \left\{ f(\boldsymbol{X}_{s_i r_i} (t_i))  \right\} + \exp \left\{ f(\boldsymbol{X}_{s_i^* r_i^*} (t_i))  \right\} } \\
    &= \log \prod_{i=1}^n \frac{ \exp \left \{ f(\boldsymbol{X}_{s_i r_i} (t_i)) - f(\boldsymbol{X}_{s_i^* r_i^*} (t_i) \right \} }{1 + \exp \left \{ f(\boldsymbol{X}_{s_i r_i} (t_i)) - f(\boldsymbol{X}_{s_i^* r_i^*} (t_i)) \right \} },
\end{align*}
where at event time $t_i$ of event $(s_i,r_i)$, a non-event $(s_i^*,r_i^*) \ne (s_i,r_i)$ is randomly sampled from the risk set $\mathcal{R}(t_i)$. The last equality corresponds to the log-likelihood of a degenerate logistic regression model with response $\boldsymbol{Y} =  (1,\dots,1)$ and non-linear predictor $\Delta_i f = f(\boldsymbol{x}_{s_i r_i} (t_i)) - f(\boldsymbol{x}_{s_i^* r_i^*} (t_i))$. 
As this partial likelihood of $N$ is an exponential dispersion family,  \cite{polinelli2024} show that, at the population level, $f_{\text{PA}}$ satisfies the following two conditions, which identify it up to a zero set:
\begin{enumerate}
    \item $f_{\text{PA}}$ maximises the population likelihood of $Y$ and $\boldsymbol{X}_{\text{PA}}$:
    \begin{equation*}
        f_{\text{PA}} = \text{arg } \underset{f}{\text{max}} \, \EX_{\boldsymbol{X}_{\text{PA}},N}{\left[ \ell (f) \right]}.
    \end{equation*}
    \item $f_{\text{PA}}$ achieves perfectly dispersed population Pearson risk:
    \begin{equation*}
        \EX_{\boldsymbol{X},N}{\left [ \frac{( Y - \dot b (\Delta_i f_{\text{PA}}) )^2}{\ddot b( \Delta_i f_{\text{PA}} )} \right ]} = a(\phi),
    \end{equation*}
    where $\dot b(\cdot)$ and $\ddot b(\cdot)$ are the first and second derivative of the cumulant generator function $b(\cdot)$, respectively.
\end{enumerate} 

Using the fact that for a binomial distribution the dispersion parameter $a(\phi)=1$ and cumulant function $b(\theta)=\log(1+\exp(\theta))$, we can now define an empirical causal discovery procedure for the causal dynamic network.

\subsection{Empirical algorithm for causal discovery in dynamic network model}
\label{subsec:empiric_alg}

An empirical version of the causal discovery algorithm can be proposed, when observing a sample $\{(\boldsymbol{x}_i(t_i), (s_i,r_i), (s_i^\ast,r_i^\ast)~|~i=1,\ldots,n\}$ of size $n$, in which the condition of having a perfectly dispersed Pearson risk is checked via a statistical test. By approximating the non-linear $f$ via a finite basis expansion $\psi$, the inference can be performed by means of generalized additive models. In particular, let $S \subseteq \{1,\dots,p\}$ a set of potential causal drivers of $N$. There exist $|S|$-dimensional basis $\psi_S$ such that $f_S:\mathbb{R}^{|S|}\rightarrow\mathbb{R}$ can be written as $f_S(\boldsymbol{x}_S) = \boldsymbol{\beta}_{S}^\top \, \psi_S (\boldsymbol{x}_S)$. In the following algorithm, the focus will be on $\boldsymbol{\beta}_S$ rather than $f_S$. 

\begin{enumerate}
    \item For any set of covariates $\boldsymbol{X}_S$, with $S \subset \{1,\dots,p\}$, find the penalised maximum likelihood estimator:
    \begin{equation*}
        \hat{\boldsymbol{\beta}}_S = \text{arg }\underset{\boldsymbol{\beta}}{\text{max}} \left [ \sum_{i=1}^n \left( \boldsymbol{\beta}^\top \psi_S (\boldsymbol{x}_{i,S}) - b( \boldsymbol{\beta}^\top \psi_S (\boldsymbol{x}_{i,S})) \right) + P_{\lambda}(\boldsymbol{\beta}) \right ],
    \end{equation*}
    where $\boldsymbol{x}_{i,S}$ is the $i^{th}$ realization of $\boldsymbol{X}_S$ and $P_{\lambda}$ is a suitable smoothness penalty function. Furthermore, we used the fact that $y_i = 1$ for all $i = 1,\dots,n$ as we are dealing with a degenerate logistic model.
    \item Find $S_1,\dots,S_K \subset \{1,\dots,p\}$ such that 
    \begin{equation*}
        H_0: \EX_{\boldsymbol{X},Y}{\left[ \frac{(Y-\dot b( \boldsymbol{\beta}^\top_{S_k} \psi_{S_k}(\boldsymbol{X}_{S_k})))^2}{\ddot b(\boldsymbol{\beta}^\top_{S_k} \psi_{S_k}(\boldsymbol{X}_{S_k}))} \right] = 1}
    \end{equation*}
    cannot be rejected. To this purpose, we use the Pearson risk statistic
    \begin{equation*}
        R(\hat{\boldsymbol{\beta}}_{S_k}) = \sum_{i=1}^n \left[ \frac{(1-\dot b(\hat{\boldsymbol{\beta}}^\top_{S_k} \psi_{S_k}(\boldsymbol{x}_{i, S_k})))^2}{\ddot b(\hat{\boldsymbol{\beta}}^\top_{S_k} \psi_{S_k}(\boldsymbol{x}_{i, S_k}))} \right],
    \end{equation*}
    in which $\hat{\boldsymbol{\beta}}_{S_k}$ denotes the penalized maximum likelihood estimator. Under the null hypothesis, the Pearson statistics is approximately chi-squared distributed, with $n - |S_k|$ degrees of freedom. Hence, a two-sided statistical test is performed, in order to select perfectly dispersed models. In practice, it has to be checked that
    \begin{equation*}
        \chi^2_{n-|S_k|,\frac{\alpha}{2}} \le \sum_{i=1}^n \left[ \frac{(1-\dot b(\hat{\boldsymbol{\beta}}^\top_{S_k} \psi_{S_k}(\boldsymbol{x}_{i, S_k})))^2}{\ddot b(\hat{\boldsymbol{\beta}}^\top_{S_k} \psi_{S_k}(\boldsymbol{x}_{i, S_k}))} \right] \le \chi^2_{n-|S_k|,1-\frac{\alpha}{2}}
    \end{equation*}
    for some significance level $\alpha$. 
    \item Among perfectly dispersed models $f_{S_1},\dots,f_{S_K}$, i.e., smooth functions associated with $\boldsymbol{\beta}_{S_1},\dots,\boldsymbol{\beta}_{S_K}$, select the one that minimises BIC as in the population case. 
\end{enumerate}

\subsection{An illustrative example}
\begin{figure}[t]
    \begin{subfigure}{0.4\textwidth}
    \raisebox{2.5cm}{
    \begin{tikzpicture}
            \Vertex[x = 0,label = X_1, Math, size = 1, color = white, fontscale = 1.5]{X1}
            \Vertex[x = 2, label = N, Math, size = 1, color = white, fontscale = 1.5]{N}
            \Vertex[x = 4, label = X_2, Math, size = 1, color = white, fontscale = 1.5]{X2}
            \Edge[Direct](X1)(N)
            \Edge[Direct](N)(X2)
        \end{tikzpicture}
        }
        \caption{}
    \label{fig:graph-2cov}% label for this sub-figure
		\end{subfigure}
    \begin{subfigure}{0.6\textwidth}
    \includegraphics[width=\linewidth]{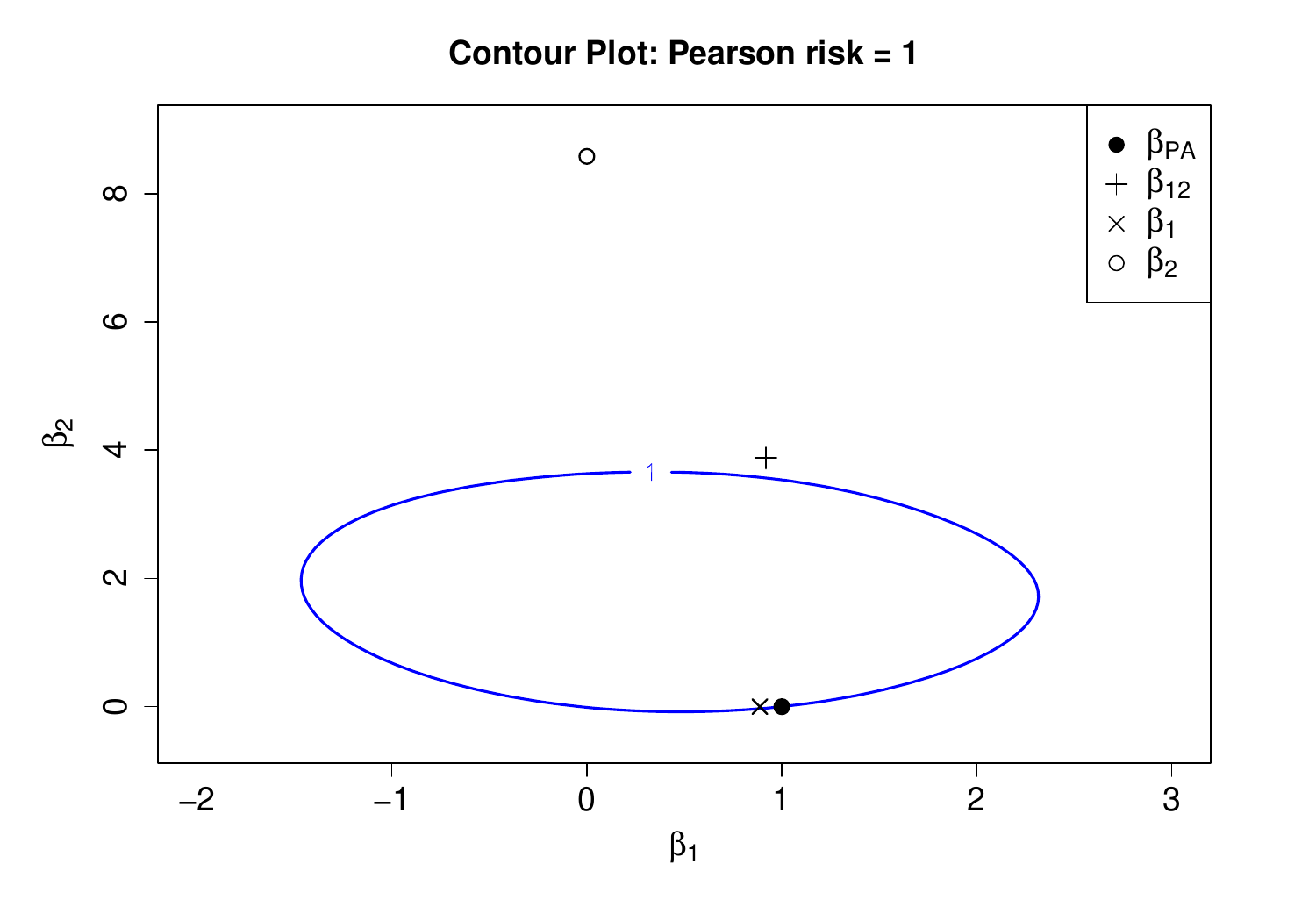}
    \caption{}
    \label{fig:contour_pr_2cov}% label for this sub-figure
     \end{subfigure}
    \caption{Structural Relational Event Process $N$ with 2 covariates $X_1$ and $X_2$.}
    \label{fig:example-2cov}% label for whole figure
\end{figure}

Let consider the structural relational event model in Figure \ref{fig:graph-2cov}. 
In this system, $X_1$ is a parent of the relational process $N$, whereas $X_2$ is its child. Hence, the generative hazard rate depends only on the first covariate. In particular, the following hazard rate has been chosen to generate data:
\begin{equation*}
    \lambda_{sr}(t) = \mathbbm{1}_{\{(s,r)\in\mathcal{R}(t)\}} \, \exp \left\{ \boldsymbol{x}_{sr, 1} (t)  \right\}.
\end{equation*}
Notice that the baseline hazard is assumed to be constant over time. Moreover,  the causal parameter vector is $\boldsymbol{\beta}_{\text{PA}}=(1,0)$. \\

In a structural relational event model with two covariates, there are three possible models to consider: (i) $X_1$ is causal parent (true model), (ii) $X_2$ is causal parent, or (iii) both $X_1$ and $X_2$ are causal parents.
Figure \ref{fig:contour_pr_2cov} shows the contour line of Pearson risk equal to 1. This clearly shows that Pearson risk invariance alone is not sufficient to identify the causal model.  Furthermore, the log-likelihood function has been maximised for all the three possible models: $\beta_{12}$ is the MLE of the full model, $\beta_1$ is the MLE of the model with only $X_1$ and $\beta_2$ is the MLE of the model with only $X_2$. The estimates are indicated by various symbols. It is clear that only $\beta_1$ lies near the contour line of the perfectly dispersed Pearson risk, close to the true causal parameter, thereby correctly identifying $X_1$ as the true causal driver. The other MLEs do not satisfy this property.

\section{Simulation study}
\label{sec:sim}
In this section, we show the effectiveness of the proposed approach by means of a simulation study. 
% It provides insights into the role the two conditions, used to define RCD, have in distinguishing between the best predictive and actual causal model.
\begin{figure}[tbp]
    \begin{subfigure}{0.49\textwidth}
    
    \resizebox{5cm}{!}{
    \begin{tikzpicture}
        \Vertex[x = 2, y = 0, label = X_1, Math, size = 1, color = white, fontscale = 1.5]{X1}
        \Vertex[x = 2, y = -2, label = X_2, Math, size = 1, color = orange!30, fontscale = 1.5]{X2}
        \Vertex[x = 4, y = -2, label = X_3, Math, size = 1, color = orange!30, fontscale = 1.5]{X3}
        \Vertex[x = 0, y = -4, label = X_4, Math, size = 1, color = white, fontscale = 1.5]{X4}
        \Vertex[x = 2, y = -4, label = N, Math, size = 1, color = white, fontscale = 1.5]{N}
        \Vertex[x = 2, y = -6, label = X_5, Math, size = 1, color = orange!30, fontscale = 1.5]{X5}
        \Vertex[x = 4, y = -6, label = X_6, Math, size = 1, color = orange!30, fontscale = 1.5]{X6}
        \Vertex[x = 6, y = -6, label = X_7, Math, size = 1, color = white, fontscale = 1.5]{X7}
        \Edge[Direct](X1)(X2)
        \Edge[Direct](X1)(X3)
        \Edge[Direct](X2)(X3)
        \Edge[Direct](X2)(X4)
        \Edge[Direct](X2)(N)
        \Edge[Direct](X3)(N)
        \Edge[Direct](N)(X5)
        \Edge[Direct](N)(X6)
        \Edge[Direct](X6)(X7)
    \end{tikzpicture}}
        \caption{}
        \label{fig:graph-7cov}% label for this sub-figure
    \end{subfigure}
    \begin{subfigure}{0.49\textwidth}
    
    \begin{equation*}
        \begin{cases}
            X_1 &= \varepsilon_1 \\
            X_2 &= X_1 + \varepsilon_2 \\
            X_3 &= X_1 - 0.5 \, X_2 + \varepsilon_3 \\
            X_4 &= X_2 + \varepsilon_4 \\
            N  &\sim \Mult \left(\mathbf{\lambda}/\Bar{\lambda} \right) \\
            X_5 &= (1-F_5) \, N + F_5 \, (1-N) + \varepsilon_5 \\
            X_6 &= (1-F_6) \, N + F_6 \, (1-N) + \varepsilon_6 \\
            X_7 &= X_6 + \varepsilon_7
        \end{cases}
    \end{equation*} 
    \caption{}
    \label{fig:sem-7cov}% label for this sub-figure
    \end{subfigure}
    \caption{Causal REM with 7 covariates with nodes in orange belonging to the Markov Blanket of the relational process $N$ with hazard rates $\lambda = \{\lambda_{sr}\}_{(s,r) \in V_1 \times V_2}$ and $\Bar{\lambda} = \sum_{(s,r)\in\mathcal{R}(t)} \lambda_{sr}(t)$. The causal parents of $N$ are $X_2$ and $X_3$.}
    \label{fig:example-7cov}
\end{figure}
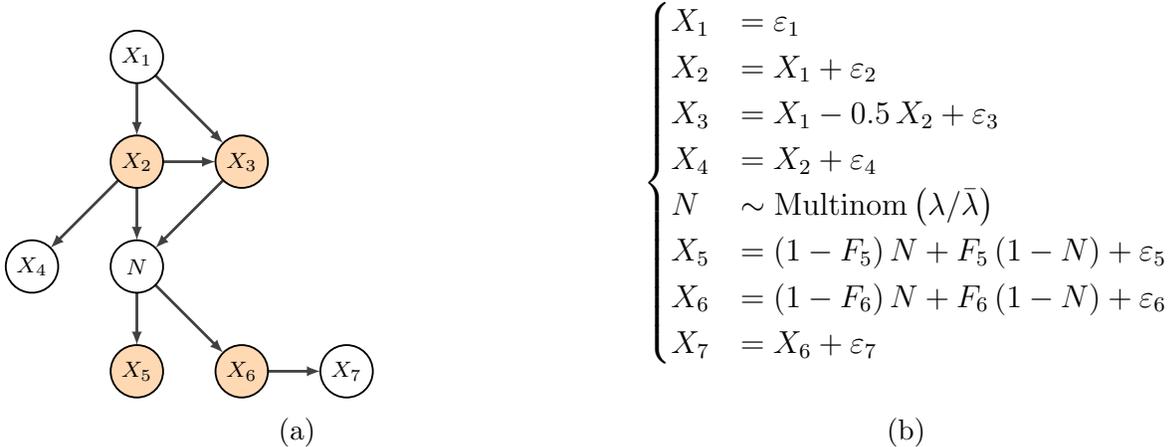
We consider a causal REM with 7 covariates as in Figure \ref{fig:graph-7cov} with data distributed according to the linear structural equation model described in \ref{fig:sem-7cov}.

The relational process $N$ has two causal parents, $X_2$ and $X_3$. We define the collection of intensity processes $\mathbf{\lambda}$, with components $\lambda_{sr}$ as in (\ref{eq:hazard-rate}), such that $f_{PA}(x_2,x_3) = 0.8 \, x_{sr, 2} (t) -0.9 \, x_{sr, 3} (t) $.
For simplicity, in this case we assume the baseline hazard to be constant over time. Furthermore we consider $V_1 = V_2 = \{1, \dots, v\}$ and that all the $v^2$ possible interactions are always at risk of happening.
Then, using a Gillespie-type algorithm \citep{gillespie1976}, we generate $n$ relational events according to this hazard. The variables $\varepsilon_i$, $ i =1,\dots 4$, are uniformly distributed over different ranges. $N$ is the multivariate counting process, handled as a matrix of dimensions $v^2 \times n$: for each column, it has an entry equal to 1 corresponding to the occurred event and 0 everywhere else. The children, namely $X_5$ and $X_6$ are obtained by flipping some entries in the columns of $N$ through the matrices $F_5$ and $F_6$ respectively. Lastly, $\varepsilon_j$ for $j = 5,6,7$ are normally distributed.

We perform 100 replications with fixed sample size $n=10^4$. As the causal REM involves 7 covariates, there are $2^7-1 = 127$ possible models. Since only linear fixed effects are assumed in this setting, each model is fitted using the R function $\texttt{glm}$. The Pearson risk is computed using the estimated parameters. 

\begin{figure}[t]
    \begin{subfigure}{0.5\textwidth}
    \includegraphics[width=\linewidth]{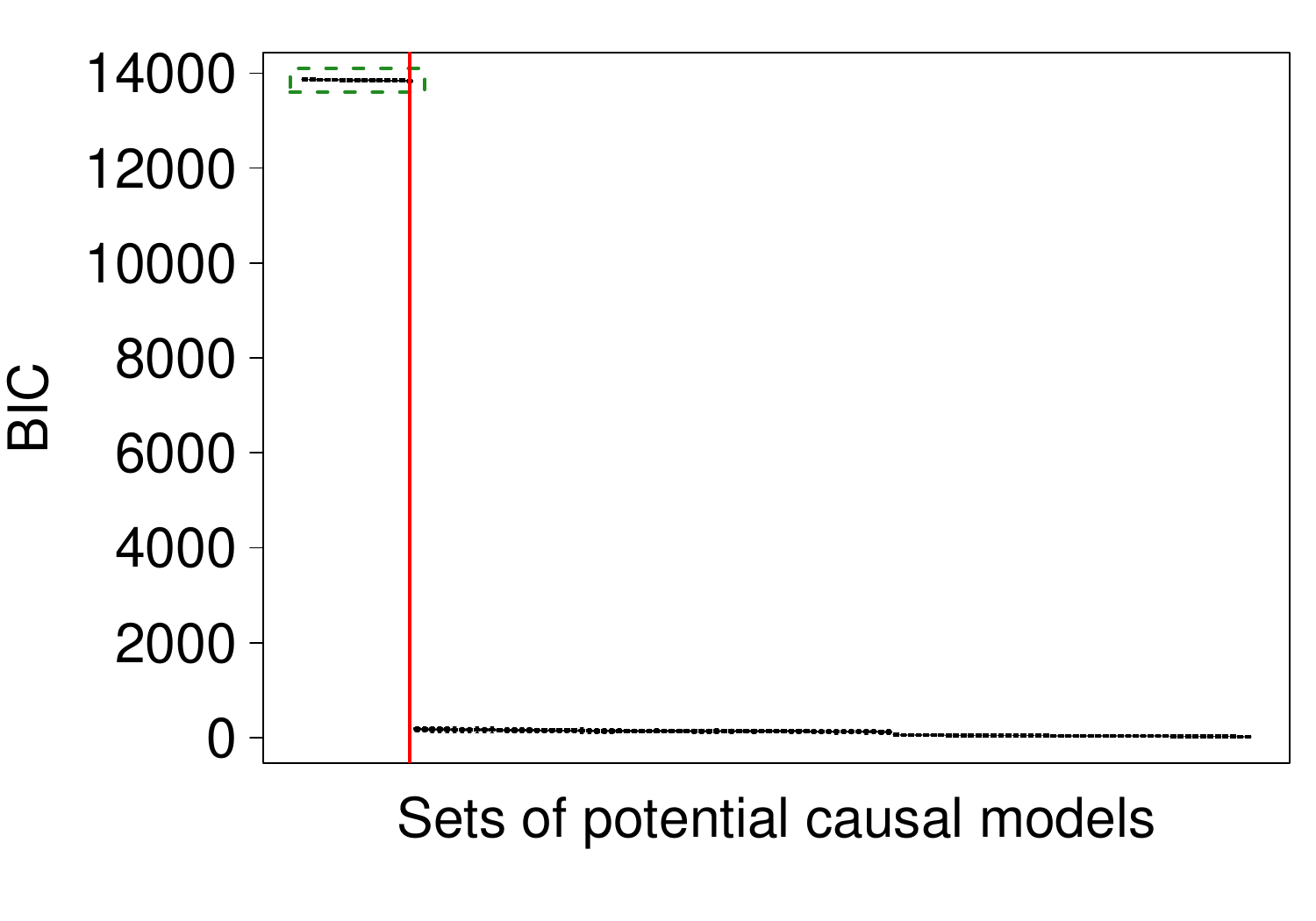}
        \caption{}
    \label{fig:bic-10k-100sim}% label for this sub-figure
				\end{subfigure}
    \begin{subfigure}{0.5\textwidth}
    \includegraphics[width=\linewidth]{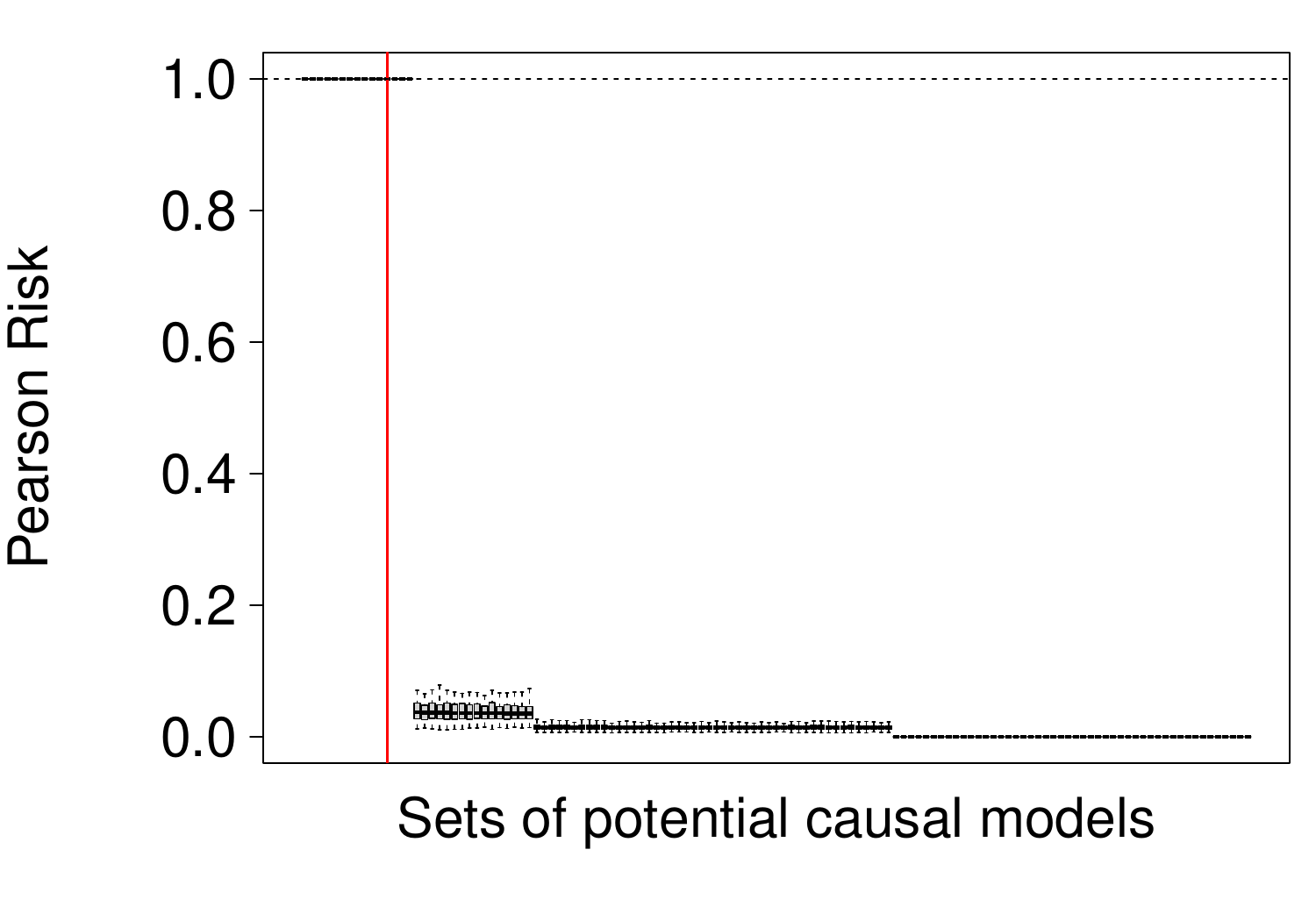}
    \caption{}
    \label{fig:pr-10k-100sim}% label for this sub-figure
    \end{subfigure}
    \caption{Results across 100 simulations. (a) The causal model (red line) is not the one that minimises BIC, but it is the one with smallest BIC among models with Pearson risk close to 1. (b)  The causal model (red line) has perfectly dispersed Pearson risk.}
     \label{fig:bic-pr-10k-100sim}
\end{figure}

\textbf{Is the true causal model predictive?} Figure \ref{fig:bic-10k-100sim} shows the boxplot of the BIC values of all the models across all the simulations: models are sorted in decreasing order according to the median of the BIC. In a purely predictive context, BIC is often used for model selection, where models with lower BIC are preferred. As expected, the true causal model (red line) is not optimal in a predictive sense having a considerably high BIC compared to other sub-models. In principle, we would expect the model with nodes belonging to the Markov Blanket of $N$ to be the most predictive and indeed in this case, the model with the lowest BIC is the one that includes only the children of the relational process, namely $X_5$ and $X_6$.

\textbf{Which models have perfectly dispersed Pearson risk?} Figure \ref{fig:pr-10k-100sim} shows the boxplot of the empirical Pearson risk for each model across all the simulations. Also in this case, values are sorted in decreasing order according to their median. The causal model (red line) has perfectly dispersed Pearson risk, but it is not the only one to satisfy this condition. Indeed, all the models containing $X_1$, $X_2$, $X_3$ and $X_4$ also have Pearson risk close to 1. Apart from the true causal parents, variables involved are either ancestors, like $X_1$, or variables blocked by causal parents, such as $X_4$. Since variables that are d-separated from $N$ in the causal graph by the causal parents are not predictive, models that contain these covariates will have greater BIC among the ones with Pearson risk close to 1. Therefore, the causal model is properly detected as the one with the lowest BIC among those with perfectly dispersed Pearson risk.

\begin{figure}[t]
    \begin{subfigure}{0.5\textwidth}
    \includegraphics[width=\linewidth]{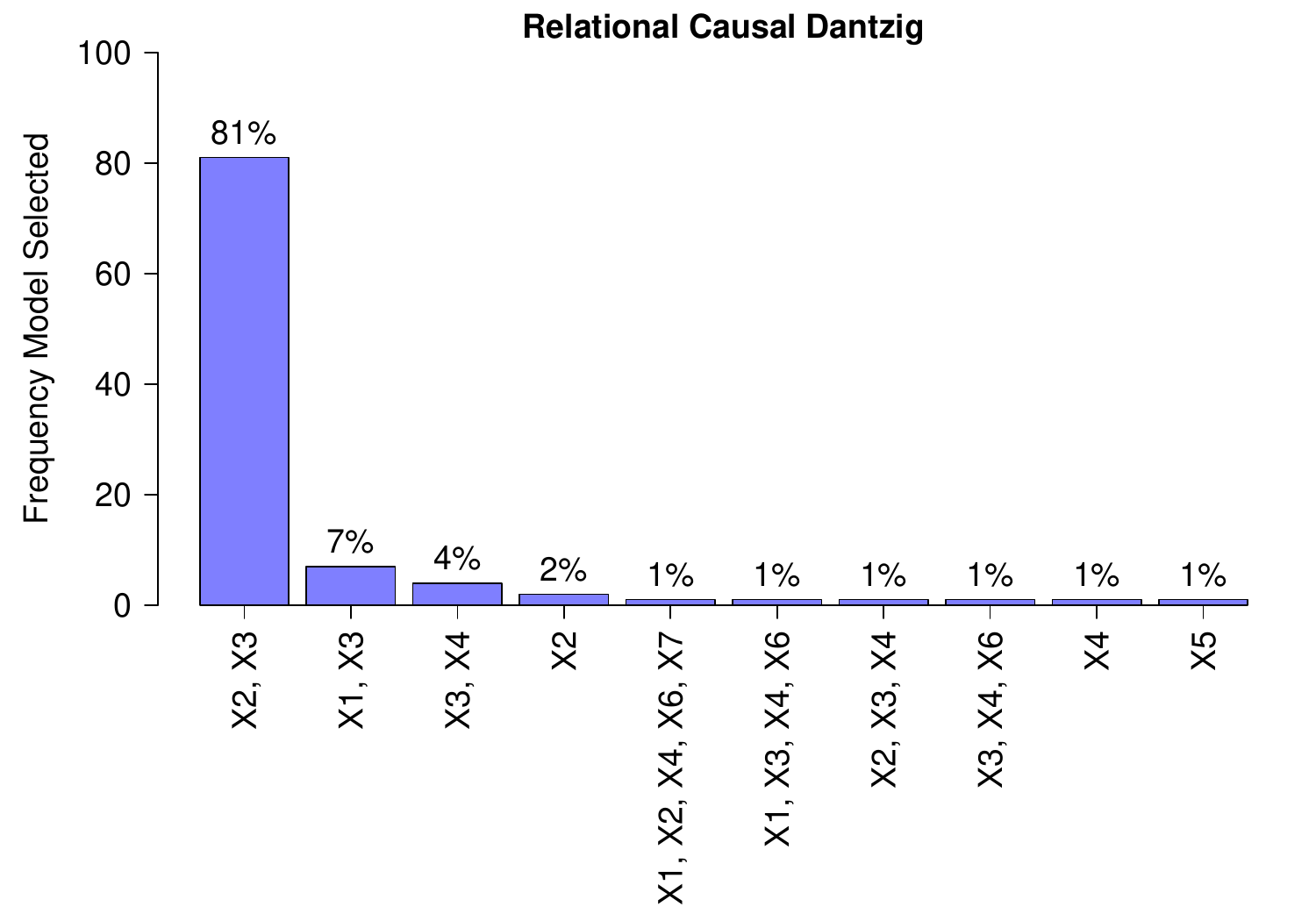}
        \caption{}
\label{fig:rcd-10k-100sim}%				
\end{subfigure}
    \begin{subfigure}{0.5\textwidth}
    \includegraphics[width=\linewidth]{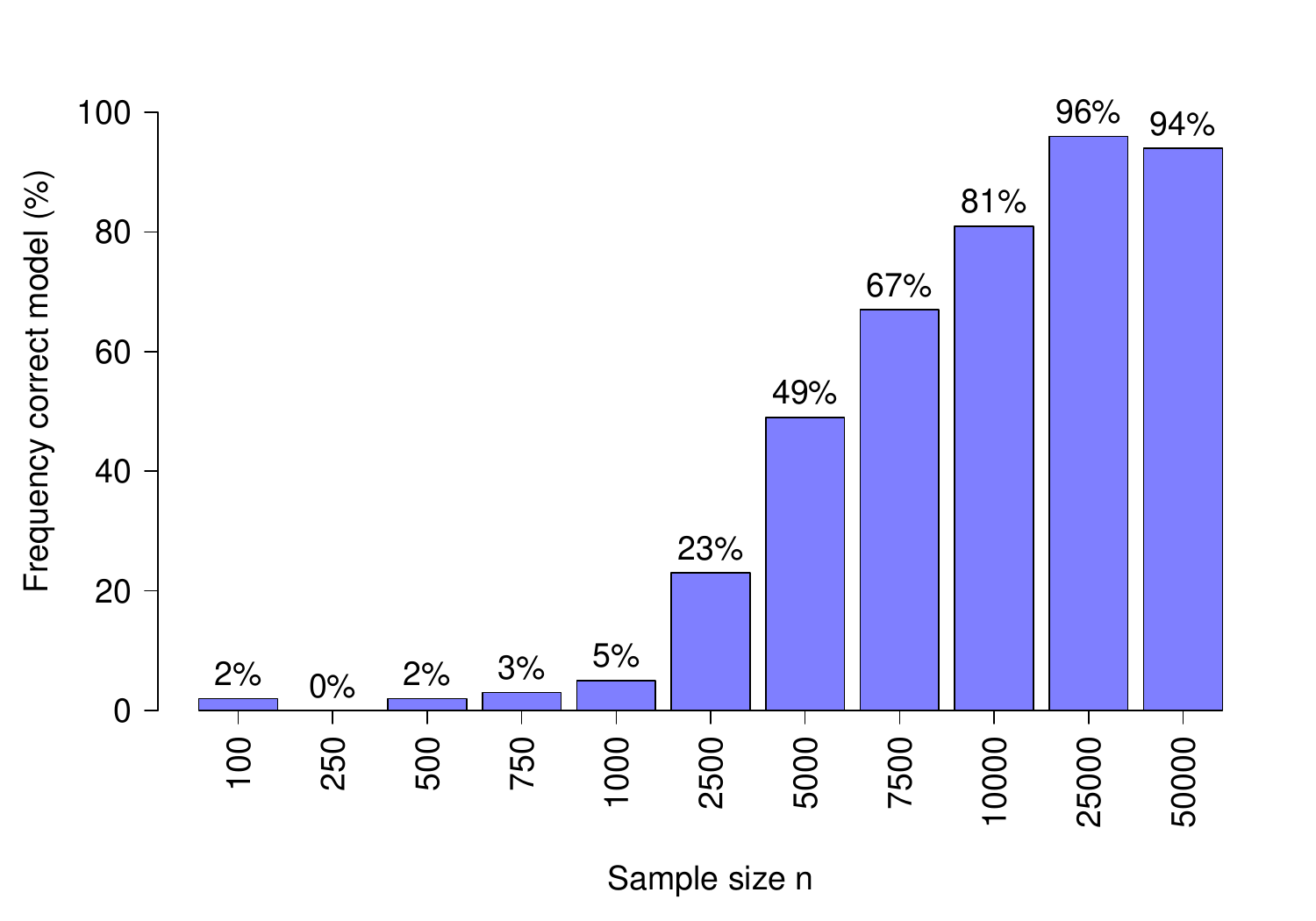}
    \caption{}
   \label{fig:freq-correct-model}% label for this sub-figure
     \end{subfigure}
    \caption{Results across 100 simulations (a) The true causal model is selected 81\% of the times for $n=10,000$. (b) The larger the sample size, the better causal discovery becomes.}
    \label{fig:hist_sim}
\end{figure}

\textbf{How often is the causal model recovered?} Figure \ref{fig:rcd-10k-100sim} exhibits all the models found by the RCD, each one with its percentage of detection, using an $\alpha=5\%$ significance level. The true causal drivers are obtained 81 times over 100 simulations.

\textbf{Does the sample size have an effect on the detection of the true causal model?} As reported in Figure \ref{fig:freq-correct-model}, the efficacy of the approach in recovering the true causal model clearly increases with the sample size $n$.

\section{Causal drivers of bike sharing}
\label{sec:bike}
In this section we provide an empirical example of the applicability of our method.
We will consider a study described in \cite{lembo2024}, which models bike sharing in Washington DC as a relational event process and investigates the effect of a variety of both global and node/edge-specific covariates. This phenomenon, clearly temporal in nature, can be represented via a dynamic network: a sequence of relational events, i.e., bike rides, between a set of vertices, i.e., bike stations, evolving over time.

\begin{figure}[t]
    \begin{subfigure}{0.5\textwidth}
    \includegraphics[width=\linewidth]{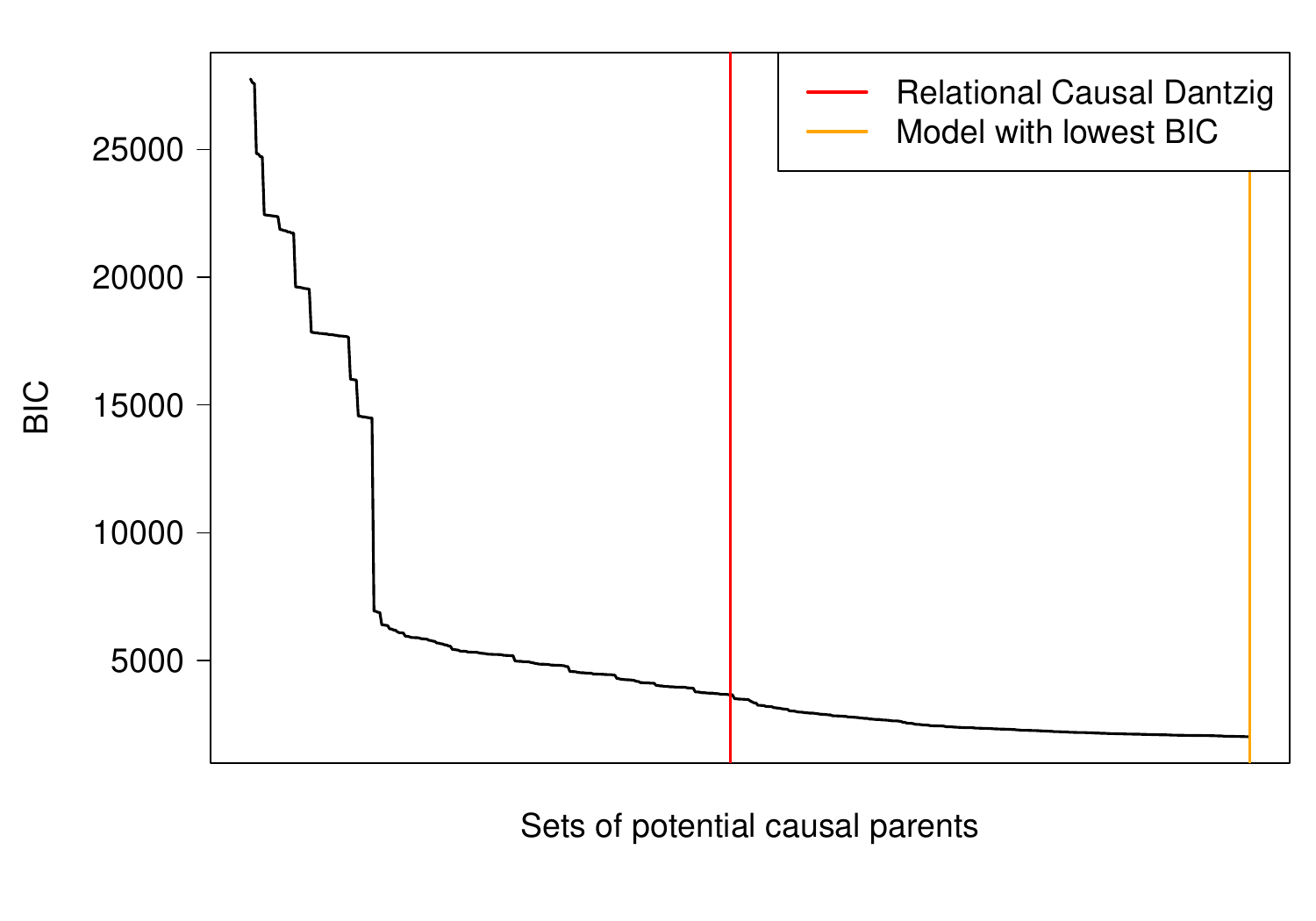}
        \caption{}
    \label{fig:bic-bike}% label for this sub-figure
		\end{subfigure}
    \begin{subfigure}{0.5\textwidth}
   \vspace{-0.5cm}
    \includegraphics[width=\linewidth]{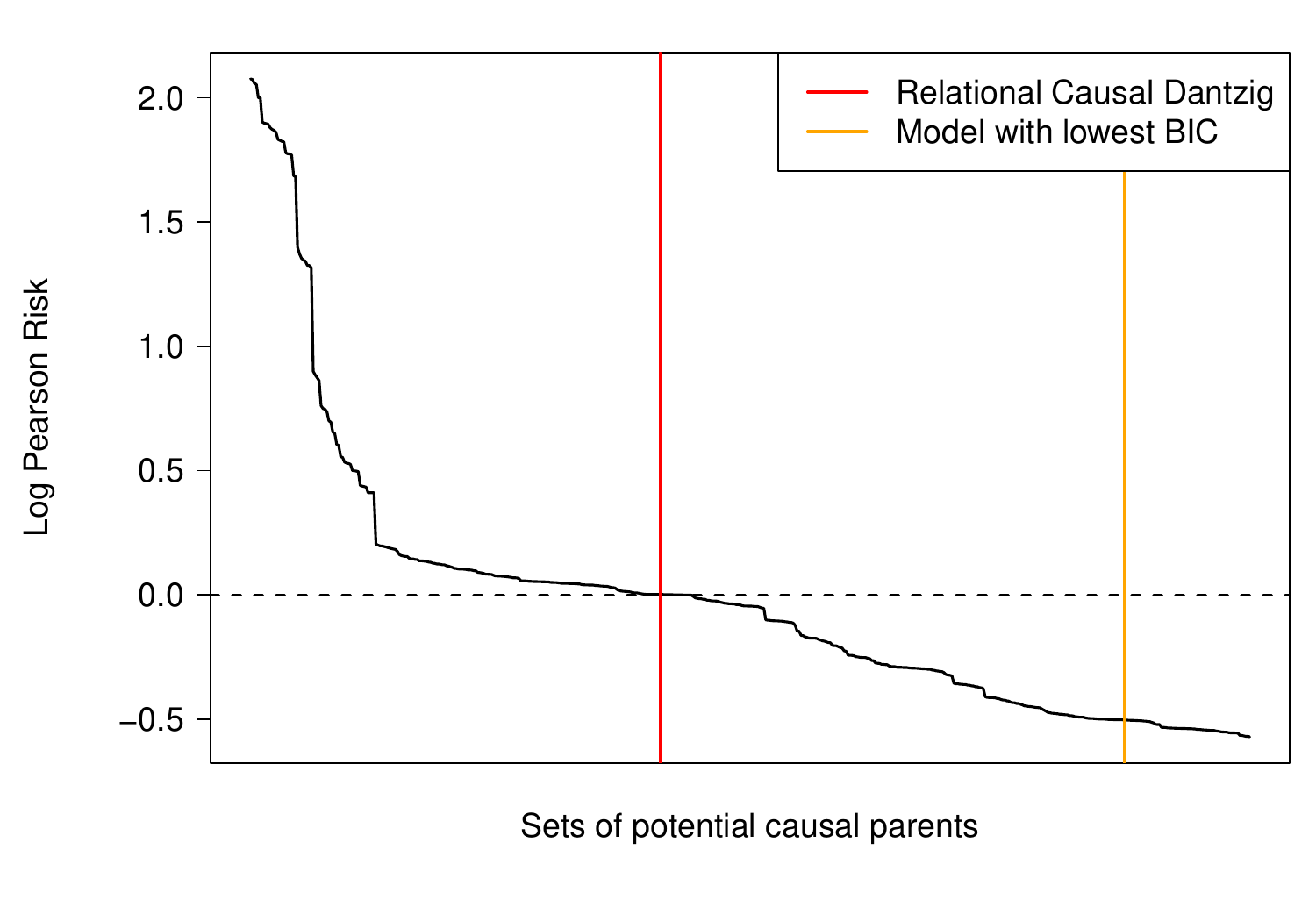}
    \caption{}
    \label{fig:pr-bike}% label for this sub-figure
     \end{subfigure}
    \caption{(a) BIC values and (b) base 10 logarithm of Pearson risk values for all models fitted on the analyzed bike sharing data. The causal model (red line) is far from being the most predictive, since it has a high BIC value but it has perfectly dispersed Pearson risk.}
    \label{fig:bic-pr-bike}% label for whole figure
\end{figure}

The data, from the bike sharing provider in Washington DC, Capital Bikeshare (\url{https://capitalbikeshare.com/system-data}), is publicly available and includes time-stamped and geo-located information on bike rides in the service area. In particular, we will consider a subset of 20,000 randomly sampled events of the dataset analyzed in that paper, which considers bike shares during the period July 9th-31st, 2023. An interaction is observed from a station $s$ to another station $r$ at time $t$ if a bike is taken from station $s$ at time $t$ and is then left at station $r$ some time after. As a result of the instantaneous nature of the events assumed in REMs, duration of the ride is ignored. 
We consider the following potential causal drivers of bike sharing:
\begin{itemize}
    \item \textbf{Global covariates}  model the role weather plays in describing this phenomena by accounting for temperature (measured in °C) and precipitation (measured in mm and log-transformed). They also account for intrinsic temporal effects like a global time effect, and a covariate for time-of-day. 
    \item \textbf{Node-level covariates}  capture any sender/receiver bike station competition effect related to their geographical position. 
    \item \textbf{Dyadic covariates} account for the distance between stations and endogenous effects of repetition, i.e., tendency to repeat the same route over time, and reciprocity, capturing the propensity to bike in along a specific route having previously bike along the same route but in the opposite direction. 
    \end{itemize}

We fit every possible sub-model (in total 511 of them, excluding the empty one) according to method proposed in \cite{lembo2024}, that extends the nested case-control partial likelihood approach, mentioned in Section \ref{subsec: icp_dyn_net}, to the case in which global covariates are of interest and also included in the hazard. We then employ the empirical algorithm described in Section \ref{subsec:empiric_alg} to recover causal drivers.    
Applying the invariant prediction procedure, results in a causal model that involves the following covariates: time of the day, sender station competition, reciprocity and repetition. Figure \ref{fig:bic-bike} shows the plot of the BIC values for each of the fitted models. The causal one (red line) is far from being the most predictive. Indeed, the model with the lowest BIC includes time of the day, sender station competition, receiver station competition, distance among stations, repetition and reciprocity. Figure \ref{fig:pr-bike} instead shows the base 10 logarithm of the Pearson risk:
the causal model is the one with the lowest BIC among those with perfectly dispersed Pearson risk.

\begin{figure}[t]
\centering
    \includegraphics[width=0.6\textwidth]{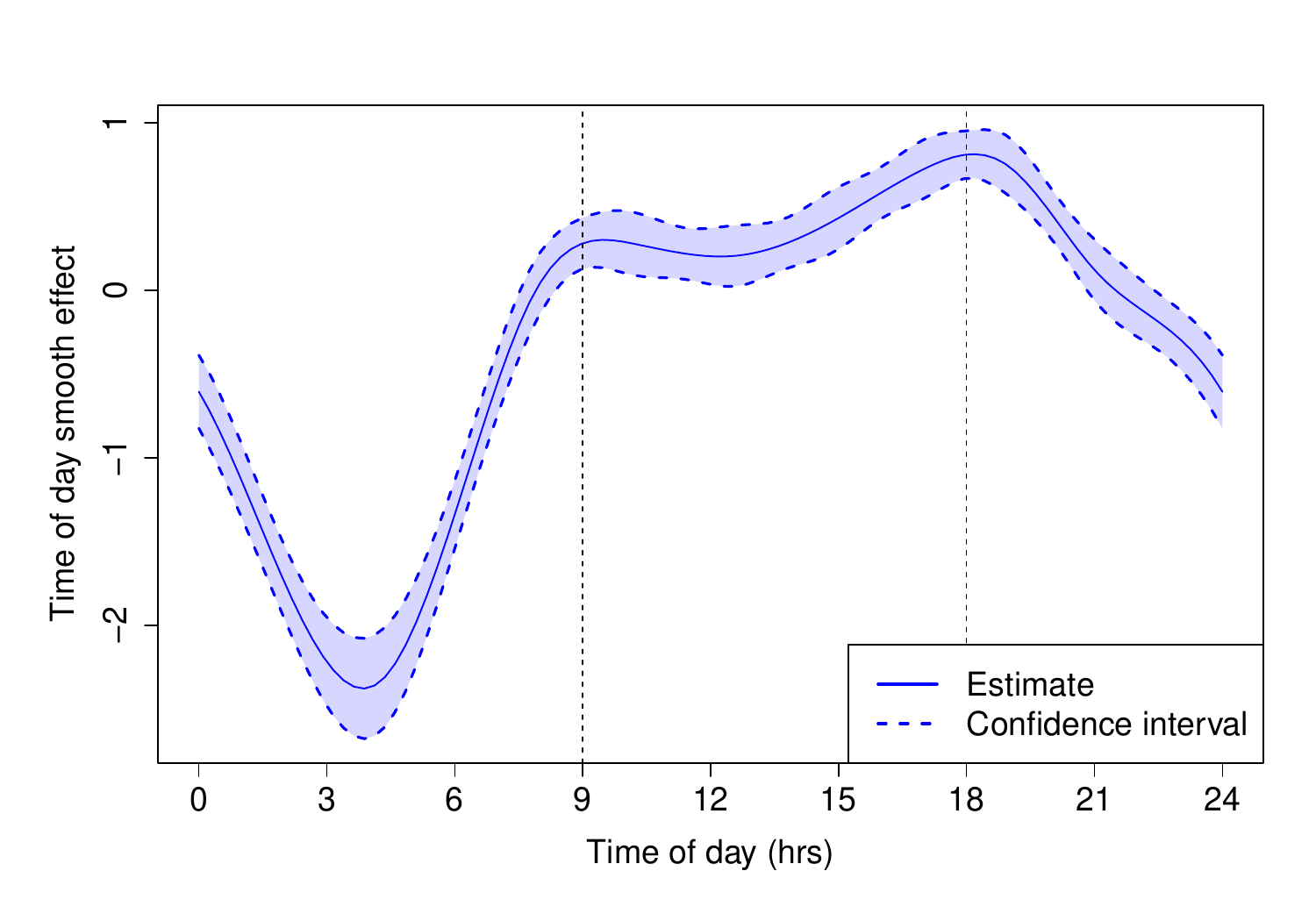}
\caption{Effect of time of the day on bike sharing. Daylight favors bike routes, with peaks at 9 a.m. and 6 p.m. when people potentially commute to work. At night, the tendency of bike sharing decreases.}% caption command
		    \label{fig:time-of-day}% label
\end{figure}

Having detected the causal drivers, we can analyze the estimated effects of these variables. Figure \ref{fig:time-of-day} shows how time of the day influences the volume of bike shares. During the night, there is a huge decline, in contrast with what happens during the day. More specifically, the plot presents two peaks at 9 a.m. and at 6 p.m., that is when workers commute to and from workplaces. Interestingly, the intensity of bike sharing is greater in the late afternoon than the peak observed in the morning.

\begin{figure}[t]
    \begin{subfigure}{0.5\textwidth}
    \includegraphics[width=\linewidth]{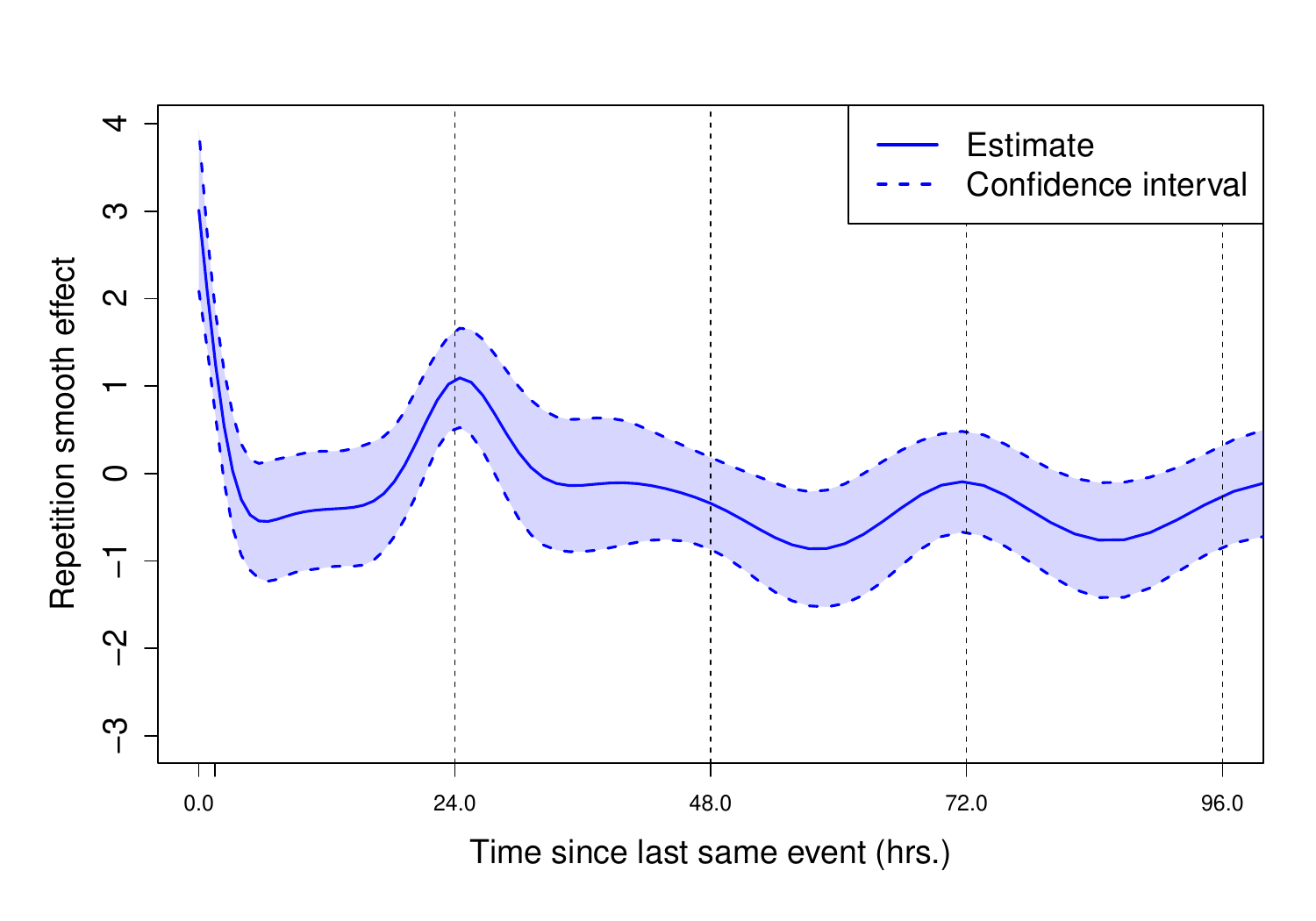}
        \caption{}
    \label{fig:rep-bike}% label for this sub-figure
		\end{subfigure}
    \begin{subfigure}{0.5\textwidth}
   \vspace{-0.5cm}
    \includegraphics[width=\linewidth]{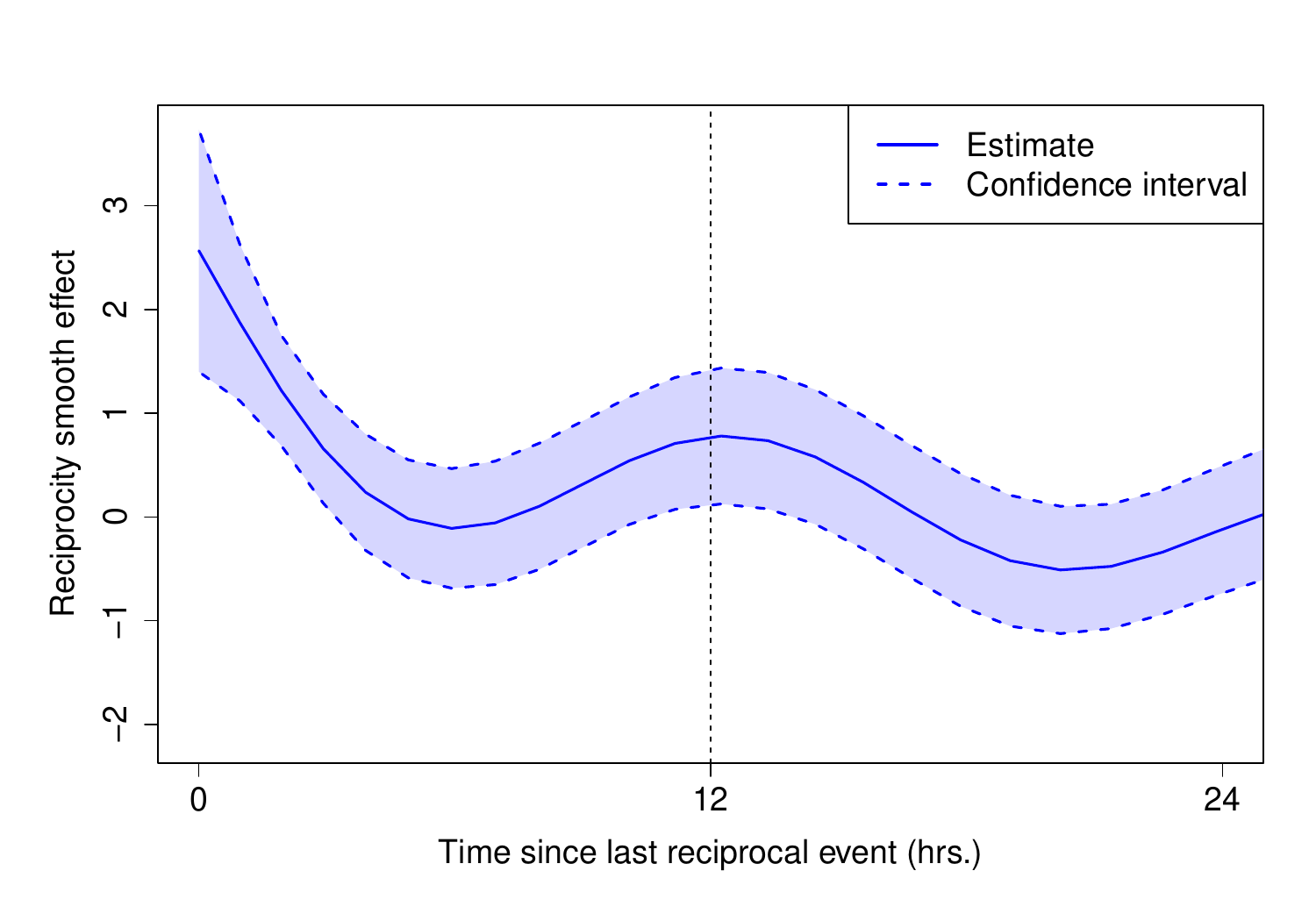}
    \caption{}
    \label{fig:rec-bike}% label for this sub-figure
     \end{subfigure}
    \caption{Dyadic endogenous effects on bike sharing: (a) Repetition captures a daily pattern suggesting a tendency to go through the same route every day. (b) Reciprocity overall generally decreases as time passes, but a peak after almost 12 hours can be noticed, potentially related to a return trip in the opposite direction.}
    \label{fig:dyadic-pr-bike}% label for whole figure
\end{figure}

From Figure \ref{fig:rep-bike}, it can be noted that there is a daily trend in the bike sharing. The peak at 24 hours suggests that the same route is traveled with a daily frequency. Lastly, Figure \ref{fig:rec-bike} reveals that, given a specific route, the tendency to observe the opposite one overall generally decreases as time passes. The peak after 12 hours could correspond to the moment in which a day work ends and people taking the same route they previously did, but in the opposite direction. 

Related to the start competition, the estimated linear fixed effect is negative, equal to -0.41270 (SE 0.0273, p-value $< 0.0001$). Interestingly, this means that there is a ``negative competition'' scenario. Indeed, the variable is defined in terms of the distance between the starting station and its closest bike station, measured in biking minutes. So, the smaller the value of the variable, the higher the competition. Thus, a negative value of the parameter could suggest that the number of stations and their placement are not enough to sustain the high demand of bike shares in the area.

\section{Conclusions}

In conclusion, this study has presented a novel approach to identifying causal drivers within dynamic network models by extending the relational event framework. Through our proposed model and empirical algorithm, we have demonstrated that invariant causal prediction is feasible with data from a single environment, which significantly enhances practical applicability in fields such as social, financial, and environmental systems. Simulation results have underscored the model’s accuracy in detecting true causal relationships. Applied to the dynamic bike-sharing network, the approach successfully revealed causal factors that align with intuitive temporal and spatial patterns in urban mobility. These findings suggest that our method provides a robust tool for causal inference in complex, temporal networks, opening avenues for further research and practical applications in data-rich, dynamic environments. Future work could explore extending this methodology to multi-layered networks and adapting it to domains with heterogeneous data structures.

\section*{Acknowledgment}
This work was supported by funding from the Swiss National Science Foundation (grant 192549).

\bibliographystyle{chicago}
\bibliography{biblio}

\end{document}